\documentclass{ws-ijmpe}

\begin{document}

\markboth{Yang Sun}{Shell model for heavy nuclei and its
application in nuclear astrophysics}

\catchline{}{}{}{}{}

\title{SHELL MODEL FOR HEAVY NUCLEI AND ITS APPLICATION IN NUCLEAR ASTROPHYSICS}

\author{\footnotesize YANG SUN}

\address{Department of Physics and Joint Institute for Nuclear
Astrophysics \\
University of Notre Dame, Notre Dame, Indiana 46556, USA\\
ysun@nd.edu}

\maketitle

\begin{history}
\received{(received date)}
\revised{(revised date)}
\end{history}

\begin{abstract}
Performing shell model calculations for heavy nuclei is a
long-standing problem in nuclear physics. The shell model
truncation in the configuration space is an unavoidable step. The
Projected Shell Model (PSM) truncates the space under the guidance
of the deformed mean-field solutions. This implies that the PSM
uses a novel and efficient way to bridge the two conventional
methods: the deformed mean-field approximations, which are widely
applied to heavy nuclei but able to describe the physics only in
the intrinsic frame, and the spherical shell model diagonalization
method, which is most fundamental but feasible only for small
systems. We discuss the basic philosophy in construction of the
PSM (or generally PSM-like) approach. Several examples from the
PSM calculations are presented. Astrophysical applications are
emphasized.
\end{abstract}

\section{Introduction}

The nuclear shell model is the most fundamental way of describing
many-nucleon systems fully quantum mechanically. However, using
the conventional shell model based on a spherical basis to study
deformed, heavy nuclei is a very difficult task because of large
dimensionality and its related problems. Even with today's
computer power, the best standard shell-model diagonalization can
be done only in the full pf-shell space, as for instance
demonstrated by Caurier {\it et al.}\cite{SM}, for which the
dimension of the configuration space may already reach one
billion. It seems impossible to apply this kind of shell model
calculation to arbitrarily large systems.

In performing calculations for a many-nucleon system, one should
choose a proper shell-model basis to start with. While in
principle, it does not matter how to prepare a model basis, it is
important in practice to use the most efficient one. In this
regard, we recognize the fact that except for a few lying in the
vicinity of shell closures, most nuclei in the nuclear chart are
deformed. This suggests strongly for shell model calculations to
use a {\it deformed} basis to incorporate correlations in large
systems efficiently. However, a deformed basis is associated with
breaking of rotational symmetry and angular momentum in that basis
is no longer a good quantum number. Although in some cases physics
may be discussed in deformed bases attached to the system
(intrinsic frames), the broken rotational symmetry should, at
least in principle, be restored by angular-momentum-projection,
which corresponds to a transformation from the intrinsic frame
back to the laboratory frame. Shell model diagonalization is then
carried out in the projected basis defined in the laboratory
frame. This is the scheme that the Projected Shell Model\cite{PSM}
is based on. It may represent a group of practical methods that
allow one to perform shell model diagonalization calculations for
deformed, heavy nuclei.

In recent years, strong demand for a shell-model treatment for
nuclei arises also from nuclear astrophysics. Since heavy elements
are made in stellar evolution and explosions, nuclear physics and,
in particular, nuclear structure far from stability, enters into
the stellar modeling in a crucial way (see the recent
review\cite{ALW05} by Aprahamian, Langanke, and Wiescher). The
nucleosynthesis and the correlated energy generation are not
completely understood, and the origin of elements in the cosmos
remains one of the unsolved physics puzzles. Nuclear shell-models
can generate well-defined wave functions, allowing one to compute,
without further approximations as often assumed in the mean-field
methods, quantities such as transition probabilities,
spectroscopic factors, and $\beta$-decay and electron-capture
rates for both ground state and excited states. These quantities
provide valuable structure information for nuclear astrophysics.
Indeed, in their example\cite{LM03}, Langanke and
Mart\'inez-Pinedo has demonstrated that shell-model calculations
could significantly modify the results of nuclear astrophysics.

\section{The projected shell model}

In performing shell-model-type calculations for large nuclear
systems, the central issue has been how to truncate the
shell-model space efficiently. It corresponds to a rearrangement
of the configuration space to decouple the most important part
from the rest of the space. There are many different schemes of
truncating a shell-model
space\cite{IBM,FDSM,PSU3,NPSM,MONSTER,MCSM}. Although these
schemes differ very much in details in the way of building the
model bases and/or choosing effective interactions, they share the
common character that the model space is first constructed by some
physical guidance. The so obtained model space contain the most
significant configurations, each of which can be a very
complicated combination in terms of the spherical shell-model
basis states. In this way, the basis dimension can be
significantly reduced and the final diagonalization is carried out
in a much smaller space, thus making a shell-model calculation for
heavy nuclei possible.

The present article reviews a particular truncation scheme
implemented in the Projected Shell Model (PSM)\cite{PSM}. One
friendly feature of the PSM is its simplicity. It starts with the
deformed Nilsson model combined by a BCS calculation to generate
quasiparticle (qp) states. The Nilsson model\cite{Nilsson} is one
of the most successful single-particle models that has been
carefully tested for different mass regions. In a PSM calculation,
the shell model truncation is first achieved within the qp states
with respect to the deformed Nilsson + BCS qp vacuum
$\left|0\right>$; then the broken rotational symmetry (also broken
particle-number and parity conservation, if necessary) is restored
for these states by standard projection techniques\cite{RS80} to
form a shell model basis in the laboratory frame; finally a
two-body Hamiltonian is diagonalized in this basis. The truncation
obtained in this way is very efficient. For example, for excited
rotational bands in a deformed, heavy nucleus (see examples
below), quite satisfactory results can be obtained by a
diagonalization within a dimension smaller than 100. In short, the
PSM starts with deformed single particle states generated by (in
principle any) mean-field solutions on one side, and performs a
shell model diagonalization like a spherical shell model (but in a
much smaller space) on the other. Such an approach lies
conceptually between the two conventional methods: deformed
mean-field model and spherical shell model, thus naturally bridges
the two methods that have coexisted but rarely have any
connections.

The Nilsson + BCS calculation defines a set of deformed qp states
(with $a^\dagger_\nu$ and $a^\dagger_\pi$ being the creation
operator for neutrons and protons, respectively) with respect to
the qp vacuum $|0\rangle$. The PSM basis is constructed in the
multi-qp states with the following forms
\begin{itemlist}
\item e-e: ~~$\{ |0\rangle, a^\dagger_\nu a^\dagger_\nu |0\rangle,
a^\dagger_\pi a^\dagger_\pi |0\rangle, a^\dagger_\nu a^\dagger_\nu
a^\dagger_\pi a^\dagger_\pi |0\rangle, a^\dagger_\nu a^\dagger_\nu
a^\dagger_\nu a^\dagger_\nu |0\rangle, a^\dagger_\pi a^\dagger_\pi
a^\dagger_\pi a^\dagger_\pi |0\rangle, \ldots \}$

\item o-o: ~~$\{ a^\dagger_\nu a^\dagger_\pi |0\rangle,
a^\dagger_\nu a^\dagger_\nu a^\dagger_\nu a^\dagger_\pi |0\rangle,
a^\dagger_\nu a^\dagger_\pi a^\dagger_\pi a^\dagger_\pi |0\rangle,
a^\dagger_\nu a^\dagger_\nu a^\dagger_\nu a^\dagger_\pi
a^\dagger_\pi a^\dagger_\pi |0\rangle, \ldots \}$

\item odd-$\nu$: ~~$\{ a^\dagger_\nu |0\rangle, a^\dagger_\nu
a^\dagger_\nu a^\dagger_\nu |0\rangle, a^\dagger_\nu a^\dagger_\pi
a^\dagger_\pi |0\rangle, a^\dagger_\nu a^\dagger_\nu a^\dagger_\nu
a^\dagger_\pi a^\dagger_\pi |0\rangle, \ldots \}$

\item odd-$\pi$: ~~$\{ a^\dagger_\pi |0\rangle, a^\dagger_\nu
a^\dagger_\nu a^\dagger_\pi |0\rangle, a^\dagger_\pi a^\dagger_\pi
a^\dagger_\pi |0\rangle, a^\dagger_\nu a^\dagger_\nu a^\dagger_\pi
a^\dagger_\pi a^\dagger_\pi |0\rangle, \ldots \}$
\end{itemlist}
for even-even, odd-odd, odd-neutron, and odd-proton nuclei,
respectively. A general term in above expressions may be written
as $|\phi_\kappa\rangle$, with $\kappa$ denoting all the quantum
numbers, such as the Nilsson quantum numbers, necessary to specify
that term. The angular-momentum-projected multi-qp states are thus
{\it building blocks} in the PSM wavefunction, which can be
generally written as
\begin{equation}
|\psi^{I,\sigma}_M\rangle=\sum_{\kappa,K\le I}
f^{I,\sigma}_{\kappa} \hat P^{\,I}_{MK}|\phi_\kappa\rangle =
\sum_{\kappa} f^{I,\sigma}_{\kappa} \hat
P^{\,I}_{MK_\kappa}|\phi_\kappa\rangle . \label{wavef}
\end{equation}
The index $\sigma$ labels states with same angular momentum and
$\kappa$ the basis states.  $\hat P^{\,I}_{MK}$ is the
angular-momentum-projection operator\cite{RS80} and the
coefficients $f^{I,\sigma}_{\kappa}$ are weights of the basis
states.

The weights $f^{I,\sigma}_{\kappa}$ are determined by
diagonalization of the Hamiltonian in the spaces spanned for
various nuclear systems as listed above, which leads to the
eigenvalue equation (for a given $I$)
\begin{equation}
\sum_{\kappa^\prime}\left(H_{\kappa\kappa^\prime}-E_\sigma
N_{\kappa\kappa^\prime}\right) f^\sigma_{\kappa^\prime} = 0.
\label{eigen}
\end{equation}
The Hamiltonian and the norm matrix elements in Eq. (\ref{eigen})
are given as
\begin{equation}
H_{\kappa\kappa^\prime}=\langle\phi_\kappa | \hat H \hat
P^I_{K_\kappa K^\prime_{\kappa^\prime}} | \phi_{\kappa^\prime}
\rangle , ~~~~~~~~~~ N_{\kappa\kappa^\prime}=\langle\phi_\kappa |
\hat P^I_{K_\kappa K^\prime_{\kappa^\prime}} |
\phi_{\kappa^\prime} \rangle . \label{z}
\end{equation}
Angular-momentum-projection on a multi-qp state
$|\phi_\kappa\rangle$ with a sequence of $I$ generates a band. One
may define the rotational energy of a band (band energy) using the
expectation values of the Hamiltonian with respect to the
projected $|\phi_\kappa\rangle$
\begin{equation}
E^I_\kappa={H_{\kappa\kappa}\over
N_{\kappa\kappa}}={{\langle\phi_\kappa | \hat H \hat P^I_{K_\kappa
K_\kappa} | \phi_\kappa \rangle}\over {\langle\phi_\kappa | \hat
P^I_{K_\kappa K_\kappa} | \phi_\kappa \rangle}} . \label{bande}
\end{equation}

The central technical issue is how to compute the matrix elements,
such as those in Eq. (\ref{z}), in the projected states. This
question applies generally to any models utilizing angular
momentum projection. Following the pioneering work of Hara and
Iwasaki\cite{HI80}, a systematic derivation has been obtained for
any one and two-body operators (of separable forces) with an
arbitrary number of quasi-particles in the projected
states\cite{PSM,PSMcode}. In principle, the projected multi-qp
basis recovers the full shell model space {\it if} all the
quasi-particles in the valence space were considered in building
the multi-qp states. However, the advantage of working with a
deformed basis is that the selection of only a few quasi-particles
near the Fermi surface is already sufficient to construct a good
shell model space. The rest can simply be truncated out.

In a usual approximation with independent quasiparticle motion
(mean-field solutions), the energy for a multi-qp state is simply
taken as the sum of those of single quasiparticles.  This is the
dominant term. The present theory modifies this quantity in the
following two steps. First, the band energy defined in Eq.
({\ref{bande}) introduces the correction brought by angular
momentum projection and the two-body interactions, which accounts
for the couplings between the rotating body and the quasiparticles
in a quantum-mechanical way. Second, the corresponding rotational
states are mixed in the subsequent procedure of solving the
eigenvalue equation ({\ref{eigen}). The energies are thus further
modified by the configuration mixing.

If the deformed states have an axial symmetry, each of the basis
states in (\ref{wavef}), the projected $|\phi_\kappa\rangle$, is a
$K$-state.  For example, an $n$-qp configuration gives rise to a
multiplet of $2^{n-1}$ states, with the total $K$ expressed by $K
= |K_1 \pm K_2 \pm \cdots \pm K_n|$, where $K_i$ is for an
individual neutron or proton.  In this case, shell model
diagonalization, i.e. solving the eigenvalue equation
(\ref{eigen}), is completely equivalent to a $K$-mixing. The
amount of the mixing can be obtained from the resulting
wavefunctions.

The above discussion is independent of the choice of the two-body
interactions in the Hamiltonian.  In practical calculations, the
PSM uses the pairing plus quadrupole-quadrupole Hamiltonian (that
has been known to be essential in nuclear structure
calculations\cite{Zuker96,HK99}) with inclusion of the
quadrupole-pairing term
\begin{equation}
\hat H = \hat H_0 - {1 \over 2} \chi \sum_\mu \hat Q^\dagger_\mu
\hat Q^{}_\mu - G_M \hat P^\dagger \hat P - G_Q \sum_\mu \hat
P^\dagger_\mu\hat P^{}_\mu . \label{hamham}
\end{equation}
The strength of the quadrupole-quadrupole force $\chi$ is
determined in such a way that it has a self-consistent relation
with the quadrupole deformation $\varepsilon_2$.  The
monopole-pairing force constants $G_M$ are
\begin{equation}
\begin{array}{c}
G_M = \left[ G_1 \mp G_2 \frac{N-Z}{A}\right] ~A^{-1} ,
\label{GMONO}
\end{array}
\end{equation}
with ``$-$" for neutrons and ``$+$" for protons, which reproduces
the observed odd--even mass differences in a given mass region if
$G_1$ and $G_2$ are properly chosen. Finally, the strength $G_Q$
for quadrupole pairing is simply assumed to be proportional to
$G_M$, with a proportionality constant in the range of 0.14 to
0.20, as commonly used in the PSM calculations\cite{PSM}.

\section{Examples}

In this section, we show several examples from the recent PSM
calculations. These include a light nucleus $^{48}$Cr for which a
full diagonalization of the pf-shell spherical shell model is
available, a well-deformed nucleus $^{178}$Hf in the heavy mass
region with the well-known 31-year isomer, and a gateway nucleus
$^{254}$No to the superheavy ``island of stability".

\subsection{$^{48}$Cr: A benchmark for models of pf-shell rotors}

$^{48}$Cr is a light nucleus for which an exact shell model
diagonalization in the pf-shell model space and a mean-field
analysis have been performed\cite{Cr48}. Though light, this
nucleus exhibits remarkable high-spin phenomena usually observed
in heavy nuclei: large deformation, typical rotational spectrum,
and the backbending phenomenon in which the regular rotational
band is disturbed by a sudden irregularity at a certain spin. In
the PSM calculations\cite{Hara99} for $^{48}$Cr, three major
shells (N = 1, 2, 3) are used for both neutrons and protons. The
shell model space is truncated at the deformation $\varepsilon_2 =
0.25$.

\begin{figure}[th]
\centerline{\psfig{file=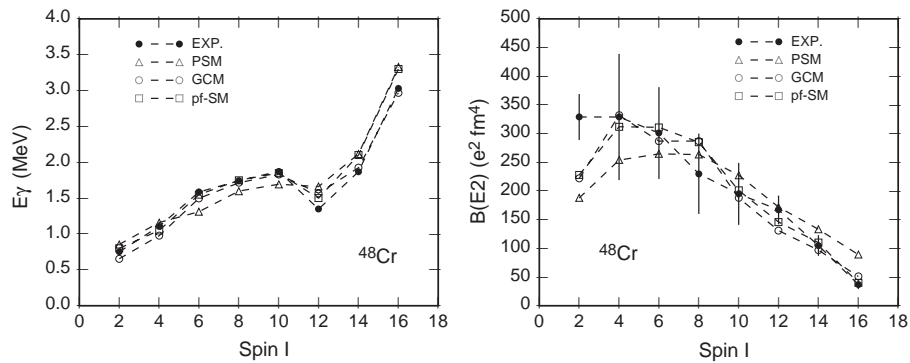,width=12cm}}
\vspace*{8pt} \caption{Left panel: $\gamma$-ray energies $E_\gamma
= E(I)-E(I-2)$ in $^{48}$Cr as functions of spin. Right panel:
B(E2) values as functions of spin. These figures are redrawn from
Figs. 1 and 2 of Ref.\protect\cite{Hara99}. }
\end{figure}

In the left panel of Fig. 1, the PSM result for $\gamma$-ray
energy along the yrast band, together with that of the pf-shell
model (pf-SM) reported in Ref.\cite{Cr48}, and that of the
Generator Coordinate Method (GCM)\cite{Hara99}, are compared with
the experimental data. One sees that the four curves are bunched
together over the entire spin region, indicating an excellent
agreement of the three calculations with each other, and with the
data. The sudden drop in E$_\gamma$ occurring around spin $I=10$
and 12 corresponds to the backbending in the yrast band of
$^{48}$Cr. In the right panel of Fig. 1, three theoretical results
for B(E2) are compared with the data. All the three calculations
use the same effective charges (0.5e for neutrons and 1.5e for
protons). Again, one sees that the theoretical descriptions agree
not only with each other but also with the data quite well. The
B(E2) values decrease monotonously after spin $I=6$ (where the
first band crossing takes place in the PSM). This implies a
monotonous decrease of the intrinsic $Q$-moment as a function of
spin, reaching finally the spherical regime at higher spins. This
implies also that the final results of a shell model calculation
do not depend on choice of the basis (either spherical
($\varepsilon_2=0$) or deformed ($\varepsilon_2=0.25$)); A
spherical shell model can describe states with large deformation
(low-spin states in Fig. 1) and a (projected) deformed shell model
can describe states having spherical properties (high-spin states
in Fig. 1).

The results shown in Fig. 1 suggest that the PSM is an efficient
shell model truncation scheme that reproduces the data with the
similar quality as the large-scale pf-SM. To compare with the
one-major-shell shell model\cite{Cr48}, the valence space in the
PSM is much larger (three major shells) yet the shell-model basis
is much smaller ($< 100$). An obvious advantage of the PSM is that
because it works in a much smaller basis with the configurations
having a clear physics meaning, the PSM is able to extract the
undergoing physics easily. In another PSM calculation\cite{Long01}
for the super-deformed band in $^{36}$Ar, the same conclusion was
drawn. The PSM reproduces the data in a much simpler way and
yields a similar quality in the results as those of the
large-scale shell model\cite{Ar36}.

\subsection{$^{178}$Hf: A deformed heavy nucleus with a long-lived isomer}

Long-lived isomers —- excited nuclear states with inhibited
electromagnetic decay —- may be considered to offer a form of
energy storage\cite{WD99,Sun05a}. The possibility to trigger the
decay by the application of external electromagnetic radiation has
attracted much interest and potentially could lead to the
controlled release of nuclear energy\cite{BS97}. The significantly
high energy of the $^{178}$Hf isomer leads to the expectation of a
lower triggering threshold, on account of the higher level
density. The high isomer energy ($\sim 2.5$ MeV) and long
half-life (31 years) in $^{178}$Hf also leads to greater potential
with regard to the utility of triggered energy release. In
determining the favorable conditions for triggering $\gamma$-ray
emission from the isomer, it is necessary to have information on
the structure of possible gateway states, as well as possible
paths of electromagnetic transitions to and from these states.
Experimentally, only a few states close to the $^{178}$Hf isomer
have been observed\cite{Hf178a,Hf178b}.

In the PSM calculation\cite{Sun04a} for $^{178}$Hf, the model
basis is built with the deformation parameters
$\varepsilon_2=0.251$ and $\varepsilon_4=0.056$. The valence space
includes three major shells (N = 4, 5, and 6 for neutrons and N =
3, 4, and 5 for protons). Fig. 2 shows the calculated energy
levels in $^{178}$Hf, compared with the known data\cite{Hf178a}.
Satisfactory agreement is achieved for most of the states, except
that for the bandhead of the first $8^-$ band and the $14^-$ band,
the theoretical values are too low. The leading structure of each
band can be read from the wave functions. We found that
\begin{itemlist}

\item the $6^+$ band has a 2-qp structure $\{ \nu [512]{5\over
2}^- \oplus \nu [514]{7\over 2}^- \}$,

\item the $16^+$ band has a 4-qp structure $\{ \nu [514]{7\over
2}^- \oplus \nu [624]{9\over 2}^+ \oplus \pi [404]{7\over 2}^+
\oplus \pi [514]{9\over 2}^- \}$,

\item the first (lower) $8^-$ band has a 2-qp structure $\{ \nu
[514]{7\over 2}^- \oplus \nu [624]{9\over 2}^+ \}$,

\item the second (higher) $8^-$ band has a 2-qp structure $\{ \pi
[404]{7\over 2}^+ \oplus \pi [514]{9\over 2}^- \}$,

\item and the $14^-$ band has a 4-qp structure $\{ \nu
[512]{5\over 2}^- \oplus \nu [514]{7\over 2}^- \oplus \pi
[404]{7\over 2}^+ \oplus \pi [514]{9\over 2}^- \}$.

\end{itemlist}
These states, together with many other states (not shown in Fig.
2) obtained from the same diagonalization, form a complete
spectrum including the high-$K$ isomeric states and candidate
gateway states.

\begin{figure}[th]
\centerline{\psfig{file=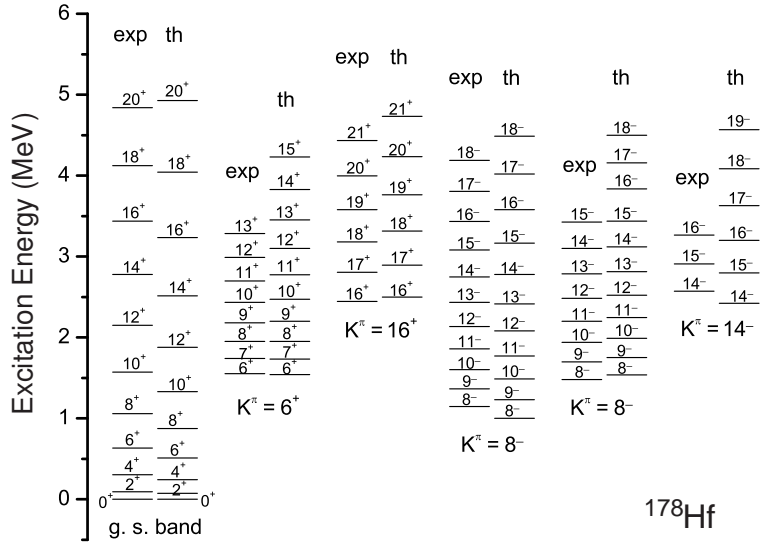,width=10cm}}
\vspace*{8pt} \caption{Comparison of calculated energy levels in
$^{178}$Hf with known data\protect\cite{Hf178a}. This figure is
adopted from Ref.\protect\cite{Sun04a}.}
\end{figure}

This example has demonstrated that the PSM can be an appropriate
nuclear structure theory for studying high-spin isomers and
associated excitations, including potential gateway states for
triggering the isomer decay. Since the diagonalization mixes the
$K$-states, the resulting wavefunctions contain valuable
information on $K$-mixing. One is thus able to use the
wavefunctions to calculate the inter-band transitions and to
analyze the degree of $K$-mixing.

\subsection{$^{254}$No: A gateway nucleus to the superheavy ``island of stability"}

One key question in our understanding of the chemical building
blocks of the Universe is what is the heaviest element that can
exist. Intimately linked to this question is the problem in
nuclear physics of the shell stabilization of superheavy nuclei.
In a simplistic model of the nucleus, the repulsive force between
more than 100 positively charged protons is sufficient to overcome
the attractive strong nuclear force and induce the nucleus to
fission. However, protons and neutrons fall into shell structures
which, in analogy to the enhanced stability of noble gases, give
extra binding to nuclei in a closed shell configuration. The
largest binding comes from a coincidence of closed shells for both
protons and neutrons, as in $^{16}$O (Z = N = 8), $^{40}$Ca (Z = N
= 20), $^{132}$Sn (Z = 50, N = 82) and $^{208}$Pb (Z = 82, N =
126).

What is the next closed shell for protons, i.e. where is a region
of long-lived, or even stable, superheavy elements beyond the
actinides? Modern theoretical calculations disagree on the size
and position of the island of stability. For the possible sources
that lead to the disagreement, one may question the
single-particle levels in those calculations. One accessible way
to gain information on these states with today's experimental
situation is to study the high-$K$ isomer states in Z $\approx$
100 nuclei. These isomer states are highly significant as their
structure is very sensitive to single-particle levels. Recently,
Xu {\it et al.}\cite{Xu04} have suggested that the existence of
the isomer states can result in increased survival probabilities
of superheavy nuclei.

The PSM is employed to study the structure of high-$K$ isomer
states in $^{254}$No and the obtained results are compared with
data from the recent experiment by Herzberg {\it et
al.}\cite{Nature06}. In the PSM calculation, three major shells
are activated for both neutrons and protons (N = 5, 6, 7 for
neutrons and 4, 5, 6 for protons). The deformed basis for the
$^{254}$No calculation is constructed with $\varepsilon_2=0.25$.
For the pairing interaction strengths in the Hamiltonian, we take
$G_1=21.24$, $G_2=13.13$, and $G_Q=0.14$.

\begin{figure}[th]
\centerline{\psfig{file=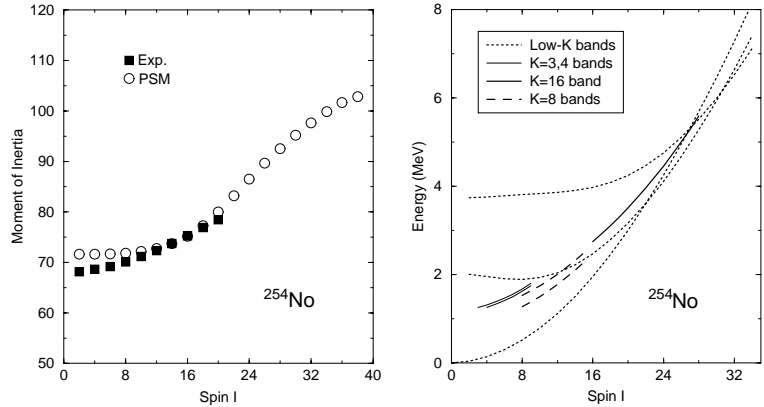,width=10cm}}
\vspace*{8pt} \caption{Left panel: Comparison of the predicted
yrast band in $^{254}$No with known data\protect\cite{Nature06}.
Right panel: Selected band energies before band-mixing, which
represent the most important configurations.}
\end{figure}

We first compare the theoretical results with the known $^{254}$No
data\cite{Nature06} for the yrast band.  As can be seen in the
left panel of Fig. 3, the theory has well reproduced the data. The
current data ends at $I=20$ where the rotational alignment process
is about to start. After that spin, the theory predicts a weak
up-bending in moment of inertia. In the right panel of Fig. 3,
band energies (as defined in Eq. (\ref{bande})) of several
important configurations are plotted to understand why this
up-bending occurs. While the 0-qp band lies low in energy till
$I=20$, the low-$K$ 2-qp band ($K^\pi=1^+, \nu [743]{7\over 2}^-
\oplus \nu [734]{9\over 2}^-$) crosses the 0-qp band at $I=22$ and
becomes the lowest one for the next spin states.  At $I=30$, the
2-qp band is crossed by the low-$K$ 4-qp band ($K^\pi=0^+, \nu
[743]{7\over 2}^- \oplus \nu [734]{9\over 2}^- \oplus \pi
[633]{7\over 2}^+ \oplus \pi [624]{9\over 2}^+$), and the latter
becomes the lowest one for the high-spin region.  The predicted
up-bending effect centering at $I=26$ in the left panel is thus
attributed to the two successive band crossings with presence of
strong band interactions.

Among the high-$K$ 2-qp states, the one with $K^\pi=8^-$ coupled
from two neutrons ($\nu [624]{7\over 2}^+ \oplus \nu [734]{9\over
2}^-$) is found to be the lowest in energy.  The next lowest $8^-$
state is from two protons ($\pi [514]{7\over 2}^- \oplus \pi
[624]{9\over 2}^+$).  These two $8^-$ bands are found
well-separated from the rest 2-qp bands.  The predicted bandhead
energies for the two $8^-$ bands are 1.266 and 1.516 MeV,
respectively, very close to the experimental $8^-$ isomer at 1.293
MeV \cite{Nature06}.

The combination of these two 2-qp $8^-$ states can give a 4-qp
state with the highest possible $K$.  This $K^\pi=16^+$ state thus
has the configuration ($\nu [624]{7\over 2}^+ \oplus \nu
[734]{9\over 2}^- \oplus \pi [514]{7\over 2}^- \oplus \pi
[624]{9\over 2}^+$). This is the lowest 4-qp band found in the
calculation, which is again well-separated (in energy and in $K$)
from other 4-qp states. Although the bandhead state ($I=16$)
already lies in a dense region, its unique $K$ quantum number
hinders its decay. The predicted bandhead energy for the $16^+$
band is 2.750 MeV. Very likely, this is the structure for the
experimental 184-$\mu$s isomer\cite{Nature06}.

\section{Application in nuclear astrophysics}

Nuclear structure input is important, and sometimes crucial, for
nuclear astrophysics. For example, an isomeric state can
communicate with its ground state through thermal excitations in
hot astrophysical environments. This could alter the elemental
abundances produced in nucleosynthesis. We are just
beginning\cite{Sun05a} to look at the impact that isomers have on
various nucleosynthesis processes such as the rapid proton capture
process thought to take place on the accretion disks of binary
neutron stars. Shell model calculations for understanding the
isomer structures and the decay rates are very much desired for
this study.

\subsection{Nuclear shape isomers in the rp-process path}

It has been suggested that in x-ray binaries, nuclei are
synthesized via the rapid proton capture process (rp
process)\cite{rp-process}, a sequence of proton captures and
$\beta$-decays responsible for the burning of hydrogen into
heavier elements. The rp process proceeds through an exotic mass
region with N $\approx$ Z, where the nuclei exhibit unusual
structure properties. One of those is the coexistence of two or
more stable shapes in a nucleus at comparable excitation energies
in nuclei with A $\approx$ 70 - 80. The nuclear shapes include,
among others, prolate and oblate deformations. In an even-even
nucleus, the lowest state with a prolate or an oblate shape has
quantum numbers $I^\pi=0^+$. An excited $0^+$ state may decay to
the ground $0^+$ state via an electric monopole (E0) transition.
For lower excitation energies, the E0 transition is usually slow,
and thus the excited $0^+$ state becomes a ``shape isomer".

\begin{figure}[th]
\centerline{\psfig{file=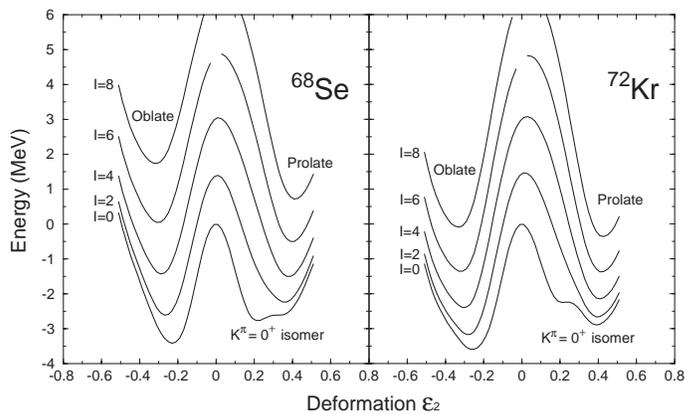,width=9cm}}
\vspace*{8pt} \caption{Energy surfaces for various spin states in
$^{68}$Se and $^{72}$Kr as a function of deformation variable
$\varepsilon_2$. This figure is adopted from
Ref.\protect\cite{Sun05b}.}
\end{figure}

Fig. 4 shows calculated total energies\cite{Sun05b} as a function
of the deformation variable $\varepsilon_2$ for different spin
states in $^{68}$Se and $^{72}$Kr. The configuration space and the
interaction strengths in the Hamiltonian can be found in the
previous calculations for the same mass region\cite{Sun04b}. Under
these calculation conditions, it is found that in both nuclei, the
ground state takes an oblate shape with $\varepsilon_2\approx
-0.25$. As spin increases, the oblate minimum moves gradually to
$\varepsilon_2\approx -0.3$. Another local minimum with a prolate
shape ($\varepsilon_2\approx 0.4$) is found to be 1.1 MeV
($^{68}$Se) and 0.7 MeV ($^{72}$Kr) high in excitation. Bouchez
{\it et al.}\cite{Kr72a} observed the 671 keV shape-isomer in
$^{72}$Kr with half-life $\tau = 38\pm 3$ ns. The one in $^{68}$Se
is our prediction\cite{Sun05b}, awaiting experimental
confirmation. Similar isomer states have also been calculated by
Kaneko, Hasegawa, and Mizusaki\cite{Kr72b}.

Since the ground states of $^{73}$Rb and $^{69}$Br are bound with
respect to the isomers in $^{72}$Kr and $^{68}$Se, proton capture
on these isomers may lead to additional strong feeding of the
$^{73}$Rb($p,\gamma$)$^{74}$Sr and $^{69}$Br($p,\gamma$)$^{70}$Kr
reactions. However, whether these branches have any significance
depends on the associated nuclear structure parameters, such as

\begin{itemlist}
\item how strong is the feeding of the isomer states?

\item what is the lifetime of the isomer with respect to
$\gamma$-decay and also to $\beta$-decay?

\item what are the lifetimes of the proton unbound $^{69}$Br and
$^{73}$Rb isotopes in comparison to the proton capture on these
states?
\end{itemlist}

The lifetime of the isomeric states must be sufficiently long to
allow proton capture to take place. No information is presently
available about the lifetime of the $^{68}$Se isomer. Based on
Hauser-Feshbach estimates\cite{rp-process} the lifetime against
proton capture is in the range of $\approx$ 100 ns to 10 $\mu$s,
depending on the density in the environment. Considering the
uncertainties in the present estimates a fair fraction may be
leaking out of the $^{68}$Se, $^{72}$Kr equilibrium abundances
towards higher masses.

\begin{figure}[th]
\centerline{\psfig{file=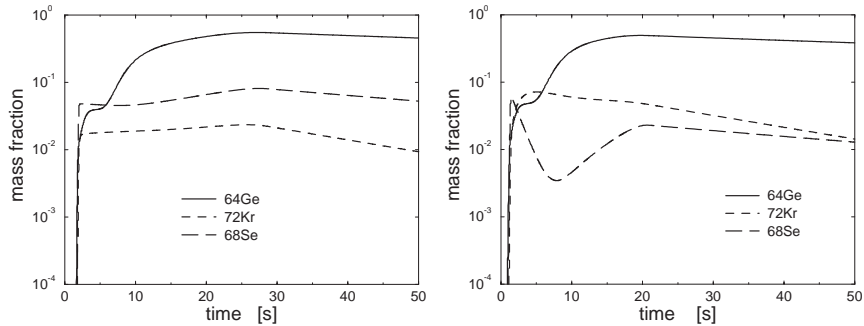,width=12cm}}
\vspace*{8pt} \caption{Mass fractions in the x-ray burst model
with two extreme cases. This figure is adopted from
Ref.\protect\cite{Sun05b}.}
\end{figure}

While it is likely that equilibrium is ensued between all these
configurations within the presently given experimental limits a
considerable flow towards higher masses through the isomer branch
cannot be excluded. Figure 5 shows the comparison between the two
extreme possibilities for the reaction sequence calculated in the
framework of a multi-mass-zone x-ray burst model\cite{Fisker05}.
The left panel shows the mass fractions of $^{64}$Ge, $^{68}$Se,
and $^{72}$Kr as a function of time neglecting any possible isomer
contribution to the flow. The right panel shows the results from
the same model assuming full reaction flow through the isomeric
states in $^{68}$Se and $^{72}$Kr rather than through the
respective ground states. The main differences in $^{68}$Se and
$^{72}$Kr mass fractions are due to rapid initial depletion in the
early cooling phase of the burst. This initial decline is
compensated subsequently by decay feeding from the long lived
$^{64}$Ge abundance. The results of our model calculations are
based on upper and lower limit assumptions about the role of the
shape isomer states. Improved calculations would require better
nuclear structure data to identify more stringent limits on the
associated reaction and decay rate predictions.

\subsection{Weak interaction rates}

The knowledge on weak interaction processes is another important
ingredient for resolving astrophysical problems. The decisive role
played by the Gamow-Teller (GT) transitions has since long been
recognized. It has been suggested that the nuclear shell model is
the most preferable method for GT transition calculations. This
has recently been emphasized by Langanke and
Mart\'inez-Pinedo\cite{LM03}.

For a theoretical model employed in GT transition calculations, it
is generally required that the model can reproduce a wide range of
structure properties of relevant nuclei. It has been shown that
the conventional shell-model diagonalization method is indeed
capable of performing such calculations. For example, Wildenthal
and Brown\cite{Wild84,BW85} obtained nuclear wave functions in the
full $sd$-shell model space, which were successfully applied to
calculation of GT rates in the $sd$ shell nuclei\cite{Oda94}.
Langanke and Mart\'inez-Pinedo\cite{LM01} made the shell-model GT
rates available also for the $pf$ shell nuclei. Still, these
sophisticated calculations are tractable only for nuclei up to the
mass-60 region, and cannot be applied to heavier nuclei which play
important roles in the nuclear processes in massive stars.

In GT transition calculations for heavy nuclei, the PSM
wavefunctions can be used. Here, we wish to list a few attractive
features in this approach, which may be relevant for future
astrophysical applications.

\begin{itemlist}

\item Because of the way the PSM constructs its basis, the
dimension of the model space is small (usually in the range of
$10^2 - 10^4$). With this size of basis, a state-by-state
evaluation of GT transition rates is computationally feasible.
This feature is important because in stellar environments with
finite temperatures, the usual situation is that the thermal
population of excited states in a parent nucleus sets up
connections to many states in a daughter by the GT operator. Our
current knowledge on GT transitions from excited nuclear states is
however very poor, and in many cases, it must rely on theoretical
calculations.

\item The PSM is a multi-shell shell model. This feature is
desired for the processes in which forbidden transitions are
dominated by the collective response of nuclei. The corresponding
calculations must involve nuclear transitions between different
harmonic-oscillator shells, and thus require a shell model working
in a multi-shell model space.

\item The existence of isomeric states in nuclei could alter
significantly the elemental abundances produced in
nucleosynthesis. There are cases in which an isomer of
sufficiently long lifetime can change the paths of reactions
taking place and lead to a different set of elemental abundances.
We have seen in above discussions that the PSM is indeed capable
of describing the detailed structure of isomeric states.

\end{itemlist}

Very recently, we have developed a method\cite{Gao06b} for
calculation of GT transition rates based on the PSM. We expect
that this method can generally be applied to the fields where weak
interaction processes take place in nuclear systems. In
particular, one may find interesting applications to cases where a
laboratory measurement for certain weak interaction rates is
difficult and where the conventional shell model calculations are
not feasible. Potential applications in nuclear astrophysics are
calculations of $\beta$-decay rates for the r-process and the
rp-process nucleosynthesis, and electron-capture rates for the
core collapse supernova modelling.

\section{Summary}

It has been desired that one enjoys advanced shell model
diagonalization methods while pushing the calculations to heavy
nuclear systems. However, most of the updated spherical shell
model calculations are severely limited in the fp-shell nuclei or
nuclei in the vicinity of shell closures. We have shown that the
Projected Shell Model may be an efficient truncation scheme for a
shell model solution. While it takes the advantages from the
mean-filed results, the PSM also contains the shell model
character that it performs configuration mixing in a truncated
basis. We have shown through the calculation for $^{48}$Cr that
the PSM results are comparable with those obtained by the
large-scale shell model calculations based on the spherical basis.
We have also presented examples for heavy and superheavy nuclei.

The use of a schematic interaction makes the theory extremely
powerful for practical applications. It is rather amazing that
such simple schematic forces can describe a large number of
different types of nuclei consistently and accurately. It may
imply that they indeed represent the most important parts of the
effective nucleon correlations in nuclei, as discussed in
Refs.\cite{Zuker96,HK99}.

An important shell model application is to provide the structure
information and decay rates for nuclear astrophysics. There have
been some shell model applications along this line (see, for
example, Ref.\cite{LM03}); however, it still has a very long way
to go. A large, unexplored region is the region away from the
$\beta$-stability, in which nuclear astrophysics is much
interested. There are many questions for every shell model,
including the PSM, to answer; the immediate ones are those
concerning the choice of proper effective interactions and single
particle states for the new regions.

\section*{Acknowledgements}

The author would like to thank the organizers of the 2006
International Conference on Nuclear Structure Physics in Shanghai
for inviting him to the conference. He is very grateful to the
following people for many discussions and contributions to this
paper: A. Aprahamian, Y.-S. Chen, J. Fisker, Z.-C. Gao, M. Guidry,
M. Hasegawa, K. Kaneko, G.-L. Long, T. Mizusaki, H. Schatz, P. M.
Walker, M. Wiescher, C.-L. Wu, F.-R. Xu, E.-G. Zhao, X.-R. Zhou,
as well as R.-D. Herzberg and the $^{254}$No experimental
collaborators. This work is supported by the National Science
Foundation of USA under contract PHY-0140324 and PHY-0216783.


\begin{thebibliography}{0}

\bibitem{SM} E. Caurier, G. Mart\'inez-Pinedo, F. Nowacki, A. Poves,
and A. P. Zuker, {\it Rev. Mod. Phys.} {\bf 77} (2005) 427.

\bibitem{PSM} K. Hara and Y. Sun, {\it Int. J. Mod. Phys.} {\bf E4}
(1995) 637.

\bibitem{ALW05} A. Aprahamian, K. Langanke, and M. Wiescher, {\it Prog.
Part. Nucl. Phys.} {\bf 54} (2005) 535.

\bibitem{LM03} K. Langanke and G. Mart\'inez-Pinedo, {\it Rev. Mod.
Phys.} {\bf 75} (2003) 819.

\bibitem{IBM} F. Iachello and A. Arima, {\it The Interacting Boson
Model} (Cambridge University Press, Cambridge, 1987).

\bibitem{FDSM} C.-L. Wu, D. H. Feng, and M. W. Guidry,
{\it Adv. Nucl. Phys.} {\bf 21} (1994) 227.

\bibitem{PSU3} G. Popa, J. G. Hirsch, and J. P. Draayer, {\it Phys.
Rev.} {\bf C62} (2000) 064313.

\bibitem{NPSM} Y. M. Zhao, N. Yoshinaga, S. Yamaji, J. Q. Chen,
and A. Arima, {\it Phys. Rev.} {\bf C62} (2000) 014304.

\bibitem{MONSTER} K. W. Schmid, {\it Prog. Part. Nucl. Phys.} {\bf 52}
(2004) 565.

\bibitem{MCSM} T. Otsuka, M. Honma, T. Mizusaki, N. Shimizu, and Y.
Utsuno, {\it Prog. Part. Nucl. Phys.} {\bf 47} (2001) 319.

\bibitem{Nilsson} S. G. Nilsson {\it et al.}, {\it Nucl. Phys.} {\bf A131} (1969) 1.

\bibitem{RS80} P. Ring and P. Schuck, {\it The Nuclear Many Body Problem}
(Springer-Verlag, New York, 1980).

\bibitem{HI80} K. Hara and S. Iwasaki, {\it Nucl. Phys.} {\bf A332} (1979) 61;
{\bf A348} (1980) 200.

\bibitem{PSMcode} Y. Sun and K. Hara, {\it Comput. Phys. Commun.} {\bf 104}
(1997) 245.

\bibitem{Zuker96}
M. Dufour and A.P. Zuker, {\it Phys. Rev.} {\bf C54} (1996) 1641.

\bibitem{HK99} M. Hasegawa and K. Kaneko, {\it Phys. Rev.} {\bf C59} (1999) 1449.

\bibitem{Cr48} E. Caurier et al., {\it Phys. Rev. Lett.} {\bf 75} (1995) 2466.

\bibitem{Hara99} K. Hara, Y. Sun, and T. Mizusaki, {\it Phys. Rev.
Lett.} {\bf 83} (1999) 1922.

\bibitem{Long01} G.-L. Long and Y. Sun, {\it Phys. Rev.} {\bf C63}
(2001) 021305(R).

\bibitem{Ar36} C. E. Svensson {\it et al.}, {\it Phys. Rev. Lett.}
{\bf 85} (2000) 2693.

\bibitem{WD99} P. M. Walker and G. D. Dracoulis, {\it Nature} {\bf 399} (1999) 35.

\bibitem{Sun05a} A. Aprahamian and Y. Sun, {\it Nat. Phys.} {\bf 1}
(2005) 81.

\bibitem{BS97} G. C. Baldwin and J.C. Solem, {\it Rev. Mod. Phys.}
{\bf 69} (1997) 1085.

\bibitem{Hf178a} S. M. Mullins {\it et al.}, {\it Phys. Lett.} {\bf B393}
(1997) 279; {\bf B400} (1997) 401.

\bibitem{Hf178b} A. B. Hayes {\it et al.}, {\it Phys. Rev. Lett.}
{\bf 89} (2002) 242501.

\bibitem{Sun04a} Y. Sun, X.-R. Zhou, G.-L. Long, E.-G. Zhao, and P. M.
Walker, {\it Phys. Lett.} {\bf B589} (2004) 83.

\bibitem{Xu04} F. R. Xu, E. G. Zhao, R. Wyss, and P. M. Walker,
{\it Phys. Rev. Lett.} {\bf 92} (2004) 252501.

\bibitem{Nature06} R.-D. Herzberg {\it et al.}, {\it
Nature} {\bf 442} (2006) 896.

\bibitem{rp-process} H. Schatz {\it et al.}, {\it
Phys. Rep.} {\bf 294} (1998) 167.

\bibitem{Sun05b} Y. Sun, M. Wiescher, A. Aprahamian, and J. Fisker,
{\it Nucl. Phys.} {\bf A758} (2005) 765.

\bibitem{Sun04b} Y. Sun, {\it Eur. Phys. J.} {\bf A20}
(2004) 133.

\bibitem{Kr72a} E. Bouchez {\it et al.}, {\it Phys. Rev. Lett.}
{\bf 90} (2003) 082502.

\bibitem{Kr72b} K. Kaneko, M. Hasegawa, and T. Mizusaki, {\it Phys. Rev.}
{\bf C70} (2004) 051301(R).

\bibitem{Fisker05} J. L. Fisker, E. Brown, M. Liebend\"orfer, H.
Schatz, and F.-K. Thielemann, {\it Nucl. Phys.} {\bf A758} (2005)
447.

\bibitem{Wild84} B. H. Wildenthal, Prog. Part. Nucl. Phys.
{\bf 11}, 5 (1984).

\bibitem{BW85}
B. A. Brown and B. H. Wildenthal, At. Data Nucl. Data Tables {\bf
33}, 347 (1985).

\bibitem{Oda94} T. Oda, M. Hino, K. Muto, M. Takahara, and K. Sato, At.
Data Nucl. Data Tables {\bf 56}, 231 (1994).

\bibitem{LM01} K. Langanke and G. Mart\'inez-Pinedo,
At. Data Nucl. Data Tables {\bf 79}, 1 (2001).

\bibitem{Gao06b} Z. C. Gao, Y. Sun, and Y. S. Chen, {\it Phys.
Rev.} {\bf C}, to be published.

\end{thebibliography}
\end{document}